\documentclass[preprint,12pt,onecolumn,letterpaper,superscriptaddress]{revtex4-2}

\usepackage{graphicx} % Include figure files
\usepackage{bm} % bold math
\usepackage{amsmath}
\usepackage{xcolor}
\usepackage{comment}

\definecolor{PRLblue}{rgb}{0.18,0.18,0.57}
\usepackage[colorlinks=true,allcolors=PRLblue]{hyperref} % add hypertext capabilities
\graphicspath{ {./}{../figures/} {figures/} {../figures/fig3_ahe_ane_plot} {../figures/fig0_overview} {../figures/fig1_expt_plot} {../figures/fig2_eng_vs_nfill/} {../figures/fig4_bands/}}
\usepackage{fancyhdr}
\begin{document}
\title{Exploring the apparent violation of the Mott relation in a noncentrosymmetric kagome ferromagnet}

\author{Benjamin Kostroun}
\affiliation{Department of Physics, Virginia Tech, Blacksburg, VA, 24061 USA}
\affiliation{Virginia Tech National Security Institute, Blacksburg, VA, 24060 USA}
\affiliation{Virginia Tech Center for Quantum Information Science and Engineering, Blacksburg, VA, 24061 USA}

\author{Tomoya Asaba}
\affiliation{Department of Physics, University of Virginia, Charlottesville, VA, 95616 USA}

\author{Sean M. Thomas} %Sean Michael Thomas
\affiliation{Materials Physics and Applications Division, Los Alamos National Laboratory, NM, 87545 USA}
%\author{Joe D. Thompson} %Joe David Thompson
%\affiliation{Materials Physics and Applications Division, Los Alamos National Laboratory, NM, 87545 USA}

\author{Eric D. Bauer} %Eric Dietzgen Bauer
\affiliation{Materials Physics and Applications Division, Los Alamos National Laboratory, NM, 87545 USA}

\author{Sergey Y. Savrasov}
\affiliation{Department of Physics, University of California, Davis, CA 95616 USA}

\author{Filip Ronning}
\affiliation{Institute for Materials Science, Los Alamos National Laboratory, NM, 87545 USA.}

\author{Vsevolod Ivanov}
\email{vivanov@vt.edu}
\affiliation{Department of Physics, Virginia Tech, Blacksburg, VA, 24061 USA}
\affiliation{Virginia Tech National Security Institute, Blacksburg, VA, 24060 USA}
\affiliation{Virginia Tech Center for Quantum Information Science and Engineering, Blacksburg, VA, 24061 USA}

\begin{abstract}
	In magnetic topological materials, time-reversal symmetry breaking gives rise to topological point and line nodes with distinctive signatures in the anomalous Hall and anomalous Nernst conductivity that satisfy the well-known Mott relation. However, this relationship can fail for doping-dependent transport measurements of
	materials with complex magnetism, topology, and electronic correlations. In this work, we present transport measurements of the correlated topological metal UCoAl doped with Ru, which appear to violate the Mott relation. We develop a model that captures the evolution of Stoner magnetism and topological Weyl points as a function of doping. Using this model, we show how the correlated flat band in this material pins the Weyl points to the Fermi energy, and demonstrate how this explains the unusual doping-dependent behavior of the anomalous Hall and anomalous Nernst conductivities in this material, while the Mott relation is in fact satisfied at each doping level. 
	%We also show a qualitative approach for understanding the behavior of doping dependent transport in UCo$_{1-x}$Ru$_{x}$Al and other topological Weyl materials.
	%
	%old abstract:
	%
	%Thermal and electronic transport measurements are a powerful tool for characterizing topological materials. In magnetic materials, time-reversal symmetry breaking gives rise to topological points and line nodes with distinctive signatures in the anomalous Hall and anomalous Nernst conductivity, leading to suggestions that doping-dependent measurements of these properties can be used to locate topological features.
	%These two anomalous transport properties arise from non-trivial Berry curvature, and necessarily satisfy the well-known Mott relation in most materials. We construct a model of a ferromagnetic Weyl metal in which the doping-dependent Stoner magnetism is treated at the mean field level. We use this model to demonstrate a ``Fermi surfing'' (FS) effect, in which the Weyls and Fermi energy shift in tandem as a function of doping. 
	%
	%We show that this FS effect leads to an apparent violation of the Mott relation, and present transport measurements of the correlated topological metal UCo$_{0.8}$Ru$_{0.2}$Al that exemplify this effect.
	%
	%This FS effect is discussed in the context of using transport measurements to locate Weyl points, and the engineering of doping-tolerant topological materials.
\end{abstract}

\maketitle

\section{Introduction}

%some intro on anomalous hall/nernst and TWM
Topological Weyl materials (TWM) have been the subject of intense study and detailed classification for nearly a decade in part due to their potential applications in photonics, thermoelectrics, and spintronics \cite{zhong2025weylsemimetalsprinciplesmaterials, 2025_thermomagnetics}. 
The topologically protected Weyl points in their electronic structures give rise to non-trivial Berry curvature (BC), and result in several characteristic properties, including surface Fermi arcs \cite{savrasov-fermi-arc}, chiral anomaly \cite{weyl-RMP}, anomalous Hall conductivity (AHC), and anomalous Nernst conductivity (ANC).
Although it is possible to directly visualize Weyl points and their surface Fermi arcs using angle-resolved photoemission spectroscopy, transport measurements of the AHC and ANC have also become critical tools for characterizing TWM. In particular, measurements of AHC and ANC as a function of doping can be used to find the energy of Weyl points relative to the Fermi level \cite{felserTaAs, noky_spectroscopy}. At low temperatures, the AHC ($\sigma_{ij}$) and ANC ($\alpha_{ij}$) are related by an energy derivative, 
%$\alpha_{ij}/T = (\pi^2 k_B^2/3e) \sigma^\prime_{ij}(\varepsilon)$ 
$\alpha_{ij}/T = (\pi^2 k_B^2/3e) (\partial \sigma_{ij}(\varepsilon)/\partial \varepsilon)$
\cite{Marder2010}, meaning either a step in the AHC or a peak in the ANC equivalently indicate the presence of a Weyl pair.

\begin{figure*}[t]
	\includegraphics[width=1.0\textwidth]{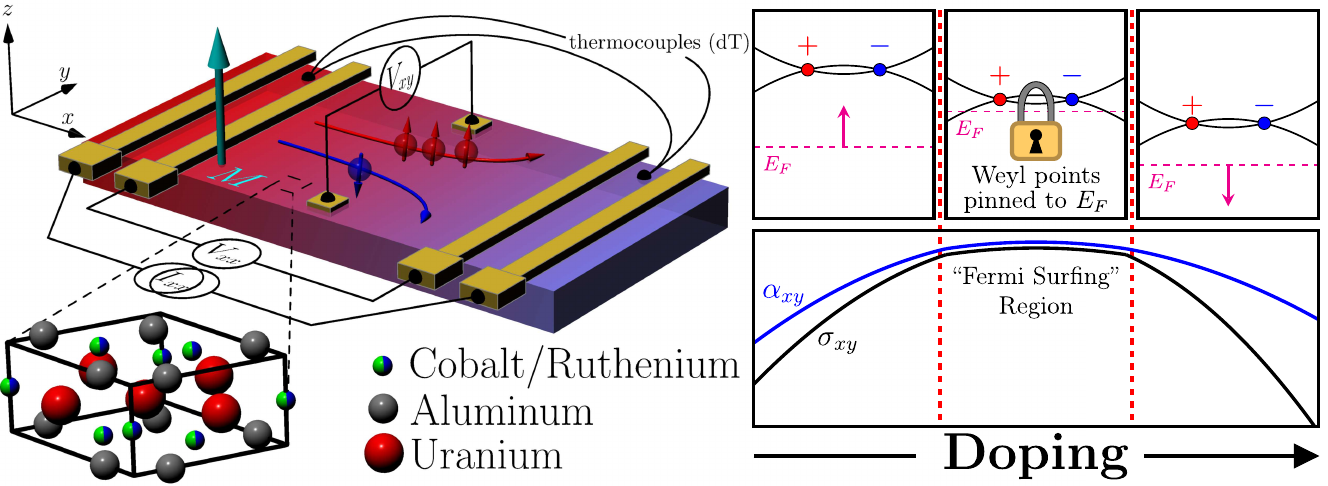}
	\caption[Overview of Experimental Setup, Structure, and Model Cartoon]{Left: overview of experimental setup used for anomalous Hall and Nernst measurements and structure of UCo$_{1-x}$Ru$_{x}$Al. Right: cartoon showing the effect of doping on transport properties; as doping increases, $E_F$ moves towards the Weyl points (red/blue circles), $E_F$ and $E_\text{Weyl}$ become pinned in the ``Fermi Surfing'' region, and then $E_F$ moves away from the Weyl points. }
	\label{overview}
\end{figure*}

%
%There have been efforts to enhance the ANC through stationary points in nodal lines, coactive-staggered features in the AHC \cite{ivanov2024coactivestaggered}, or strain tuning of the anomalous Hall and Nernst angles \cite{Meng2025_AHE_ANE_angles}. 

%explain history of the material. 2-3 paragraphs
%explain issue with mott relation
%The paper is organized as follows ...

Here, we report doping-dependent measurements of the AHC and ANC of correlated TWM UCo$_{1-x}$Ru$_{x}$Al. Both the AHC and ANC peak at the same doping level, appearing to violate the Mott relation. We assert that this apparent violation is a consequence of the doping-dependent magnetism and correlated flat bands which lead to a ``Fermi surfing'' effect, wherein the Weyl points and Fermi energy shift in tandem with doping. We present a qualitative argument showing how the doping dependence of the AHC and ANC is entirely determined by the relative motion of the Weyl points and Fermi energy. To explicitly demonstrate the origin of this effect, we develop a Weyl semimetal tight-binding model with a Hubbard interaction and study the behavior of its AHC and  ANC as a function of electronic filling. By tuning the model parameters to match a pair of critically tilted Weyl points believed to be responsible for the colossal ANC in UCo$_{1-x}$Ru$_{x}$Al \cite{ucra-sciadv}, we show that AHC and ANC can peak at the same doping level, matching our experimental observation. While this appears to violate the Mott relation, we show that the relation correctly holds as a function of energy at each doping level. This \textit{apparent} violation of the Mott relation arises due to the Fermi level being pinned to the narrow correlated bands forming the Weyl points, meaning the shapes of the AHC and ANC signals as a function of electronic filling are completely determined by the tandem motion of the Weyl points and Fermi level. 
%We also present a qualitative argument for interpreting the doping dependence of the AHC and ANC, and apply it to UCo$_{1-x}$Ru$_{x}$Al and other TWM, demonstrating the efficacy of this approach as a kind of spectroscopy to track topological features.
%Interpreting the doping dependence of the AHC and ANC in this way also has broad implications for using these transport measurements as a kind of spectroscopy to locate Weyl points in other materials. 
We also discuss how the relative motion of Weyl points and Fermi level can be used to interpret doping-dependent measurements in other materials, and how the ``Fermi surfing'' effect can be used strategy for engineering topological devices that maintain performance even when the material is doped or vacancies are created.

\begin{figure*}[t]
	%\captionsetup{justification=raggedright}
	%\includegraphics[width=1.0\columnwidth]{fig1_expt.png}
	\includegraphics[width=1.0\textwidth]{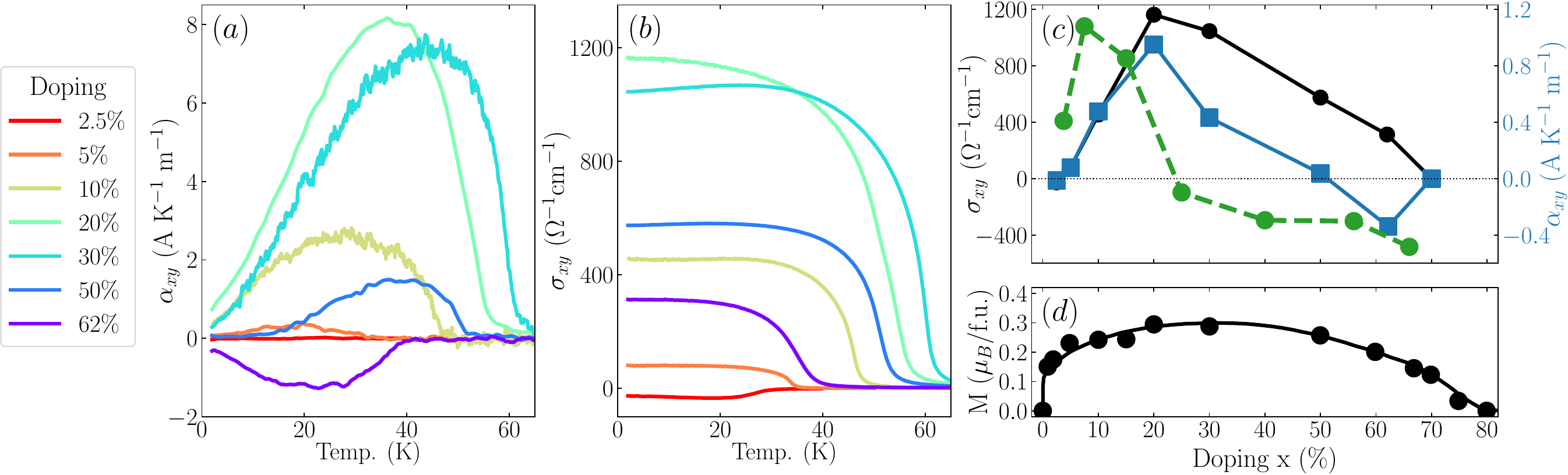}
	\caption[Magnetic and transport of UCo$_{1-x}$Ru$_{x}$Al]{(a) Measured anomalous Nernst coefficient $\alpha_{xy}(T)$ and (b) anomalous Hall conductivity $\sigma_{xy}(T)$ of UCo$_{1-x}$Ru$_{x}$Al below 65 K. (c) $\sigma_{xy}$ and $\alpha_{xy}$ as a function of Ru concentration at 3K, peaking at $x_\text{Ru}=20$\%. Dashed green line shows $\alpha_{xy}$ expected from the Mott relation using computed density of states. (d) Magnetization as a function of doping, reproduced from Ref. \cite{Andreev1996_onset_of_ferromagnetism}. }
	%\caption[Magnetic and transport of UCo$_{1-x}$Ru$_{x}$Al]{Measured anomalous Nernst coefficient $\alpha_{xy}$ of UCo$_{1-x}$Ru$_{x}$Al (a) for different dopings as a function of temperature. Anomalous Hall coefficient $\sigma_{xy}$ (b) for different dopings as a function of temperature. $\sigma_{xy}$ and $\alpha_{xy}$ as a function of Ru concentration at 3K (c). In (c), both $\sigma_{xy}$ and $\alpha_{xy}$ peak at $x_\text{Ru}=20$\%; Dashed green line (a.u.) shows ANC expected from the Mott relation assuming constant density of states. Magnetization (d) as a function of doping is reproduced from Ref. \cite{Andreev1996_onset_of_ferromagnetism}. }
	\label{expt}
\end{figure*}

UCo$_{1-x}$Ru$_{x}$Al crystallizes in the inversion broken ZrNiAl-type structure \cite{mmm} with the uranium atoms forming a distorted kagome sublattice \cite{ucra-sciadv}. At finite doping, the system exhibits itinerant ferromagnetism \cite{Andreev1996_onset_of_ferromagnetism} even though the UCoAl and URuAl endpoints are both paramagnetic \cite{Sechovsky1986_UCoAl_metamagnet} (Fig. \ref{expt}(d)). In fact, the parent compound UCoAl is considered to be a metamagnet \cite{Havela1997_metamagnet}, with even modest doping \cite{Andreev1997_UCoAl_Ru_doping}, pressure \cite{Opletal2017}, or external magnetic fields inducing a ferromagnetic state \cite{Sechovsky1986_UCoAl_metamagnet}.

%add point about orbital magnetism? \cite{2018_UCoAl_orbital}

%explain magnetism in the material - discuss how exchange interaction drives the magnetism, compare with U/J sim results
%explain width of U5f/6d bands and (that plays into the model later)
%explain how the fermi level is dragged along with the U-5f bands, and illustrate how the Weyl points will "chase" the Fermi level as the material is doped. 
The ferromagnetic state emerges at small Ru doping due to the hybridization of U-$5f$ and Ru-$4d$ states which increases the in-plane ferromagnetic coupling \cite{UCoRuAl_tricritical}. At high Ru doping, the magnetic behavior is markedly different, with the paramagnetic state of URuAl emerging in part due to the cancellation of spin $\mu_S$ and orbital $\mu_L$ moments \cite{urual_cancellation, Pospisil2016_properties_collapse}. First-principles calculations reveal that exchange interactions can explain the emergence of the ferromagnetic phase \cite{supplement}; this picture also has good agreement with the experimentally measured ratio $|\mu_L^U/\mu_S^U|$ of orbital/spin moments \cite{2018_UCoAl_orbital, 2020_co2mnga_wiedemann_franz}.

% This ferromagnetism is driven by hybridization of the U 5f and Co 3d bands
%The kagome structure and strong correlations result in a large DOS from the U-5f bands. As the material is doped, magnetism changes (cite), and the weyl points may move, though the Fermi level should remain pinned to the position of the U-5f bands due to the large DOS
%This makes this system ideal for exploring the interplay of magnetism, Weyl physics, etc. 
The ferromagnetic order, large spin-orbit coupling, and kagome flat bands \cite{Yin2022_kagome_magnets} that are further renormalized by moderately strong electronic correlations together result in UCo$_{0.8}$Ru$_{0.2}$Al having the second highest ANC of $\alpha_{xy} = 15$ A K$^{-1}$ m$^{-1}$ \cite{Heremans_MnBi} and the highest anomalous Nernst thermopower of 23 \textmu V/K \cite{ucra-sciadv}, out of all known materials. This colossal thermal transport is believed to have an intrinsic topological origin, as first principles calculations have identified over one hundred Weyl points and several nodal lines near the Fermi energy. The uranium atoms form a kagome sublattice, resulting in a set of flat, correlated U-$5f$ bands near $E_F$ that are responsible for the large number of crossings. In fact, the ANC can be explained by a single set of critically tilted Weyls separated by 0.06 1/\AA ~along $k_z$ which appear 26 meV above the $E_F$ in the 20\% doped material.
%The origin of the AHC/ANC has been attributed to the large number of topological features in the BZ, which are sources of non-trivial Berry curvature, driven by the
%in fact, the ANC can be attributed to a single set of weyl points separated by 0.06 along the kz direction that appear that appear 26 meV above the Fermi level 
Along with the complex ferromagnetism, these aspects make this system ideal for exploring the interplay of spin-orbit interactions, magnetism, Weyl physics, and strong correlations.
%alpha/sigma ratio should be constant? \cite{2020_co2mnga_wiedemann_franz}
%mention that ucoal exceeds this?

%%%%%%%%%%%%%%%%%%%%
%%  EXPT METHODS  %%
%%%%%%%%%%%%%%%%%%%%

Doping-dependent measurements of the AHC and ANC have previously been suggested as a way to track the position of Weyl points in TWM \cite{noky_spectroscopy, felserTaAs}. This approach has been used to study the interplay of magnetism and topology in a variety of TWM, including Mn$_3$Sn \cite{Yano2024_mn3sn_doping}, Mn$_3$Ge \cite{Rai_2024_Mn3Ge_doping}, Co$_2$MnAl \cite{Sakuraba2020_co2mnal_doping}, and Co$_3$Sn$_2$S$_2$  \cite{Thakur2020_co3sn2s2_ni_doping, Liu_2023_co3sn2s2_both_doping, 2020Shen_co3sn2s2_ni_doping, 2020_Shen_co3sn2s2_fe_doped}. The large magnitude of the AHC and ANC in UCo$_{1-x}$Ru$_{x}$Al make it a convenient system to use doping-dependent transport measurements to locate the aforementioned critically tilted Weyl points.

\section{Results}

The results of doping-dependent measurements (see Methods) of the AHC and ANC of UCo$_{1-x}$Ru$_{x}$Al are shown in Figure \ref{expt}. At each doping the anomalous Nernst coefficient $\alpha_{xy}$ (Fig. \ref{expt}(a)) increases from zero before a sharp dropoff at $T_C$. The maximum value of $\alpha_{xy}$ increases up to 20\% doping, gradually decreasing until it flips sign at $x=62\%$ near the quantum critical point \cite{2018_UCoAl_orbital}. The temperature at which $\alpha_{xy}$ is maximum also increases with Ru concentration up to 30\%, which may be due to an intrinsic mechanism, such as a broadening Berry curvature peak above $E_F$, or extrinsic effects due to the doping disorder. The anomalous Hall coefficient $\sigma_{xy}$ has a much simpler behavior, increasing up to 20\% doping and then decreasing monotonically until the magnetism vanishes at the ferromagnetic endpoint.

The doping-dependent behaviors of $\sigma_{xy}$ and $\alpha_{xy}/T$ at 3K are compared in Fig. \ref{expt}(c). 
%The AHC and ANC are plotted as a function of doping in Fig. \ref{expt}c, with both reaching a maximum value at 20\% Ru doping and decreasing with further doping, vanishing completely at 70\%, the upper ferromagnetic endpoint. 
The dashed line indicates the numerical derivative of the AHC with respect to doping, $(\partial \sigma_{xy}/\partial n)$. Based on the Mott relation, $\alpha_{xy}/T \propto \partial \sigma_{xy}/\partial \varepsilon = (\partial \sigma_{xy}/\partial n) (\partial n/\partial \varepsilon)$, we can compute the expected behavior of $\alpha_{xy}/T$ using the density of states obtained from first principles \cite{supplement}, which 
%this should follow the expected behavior of $\alpha_{xy}/T$, 
%assuming a constant density density of states $\partial n/\partial \varepsilon \sim \rm{const.}$, yet it 
shows a significant deviation from the measured ANC. The peak ANC occurs at a higher doping and above 25\% Ru doping the numerical derivative becomes negative, whereas the measured ANC remains positive until $x \sim 50$\%. 
While the computed density of states may differ from the true experimental value,
%The large density of states near the Fermi level \cite{ucra-sciadv}, means there will be little change as the material is doped, making $\partial n/\partial \varepsilon \sim \rm{const.}$ a reasonable assumption. 
%The Mott relation of course relates the derivative of the AHC with respect to energy, not doping, so the effect of the density of states (DOS) must be taken into consideration. 
%Nevertheless, 
even large variations in the DOS would only scale the plot along the vertical axis, and cannot explain the observed sign discrepancy. This means that this apparent violation of the Mott relationship is robust, in the sense that it cannot be explained within the framework of the rigid band approximation.

% Doping dependent measurements of AHC and ANC have been proposed as a means to locate Weyl points in topological materials, and the large magnitude of these signals in UcoRuAl makes this a convenient way to track Weyl motion in this system.

%back of the envelope derivation of dE_fermi/dx ~= dE_Weyl/dx. 

% Weyl points create signatures AHE ANE -> peaked at the Weyl energy, so it can be found
% Furthermore, they are related through mott relation, so you can have a sanity check
% in a real material, one can get energy resolution by doping, which is not the exactly same thing
% What is the effect of this doping? Well, instead we have d\sigma/dE = d\sigma/dn * dn/dE, so the derivative of the doping dependent AHE can be compared to the ANE directly, after an appropriate rescaling by the density of states. 
% For a material with many states near E_F, doping will only lead to a small change in energy, so the density of states will remain approximately constant across this range. Therefore, 
% introduce "Fermi-Surfing" idea
% 
We propose that the unusual behavior of the AHC and ANC in UCoAl can be explained through an effect we have termed ``Fermi surfing,'' stemming from the strongly correlated U-$5f$ flat bands in UCo$_{1-x}$Ru$_{x}$Al. Before presenting a detailed model, we first illustrate the essence of this effect, depicted in Figure \ref{overview}. In a magnetic material, untilted Type-I Weyl nodes will only move in $\bm{k}$-space as the magnetism is varied, remaining at the same energy. The same is not true for tilted Weyls, which will shift in both $\bm{k}$-space and energy when the magnetic band splitting is changed. In fact, critically tilted, Type-1.5 Weyls, arising from the crossing of a dispersing band and a flat band, have their energy set by the flat band. Furthermore, the large density of states of the flat band means the Fermi energy, $E_F$, will be pinned to the flat band energy. If the position of the flat band changes due to doping-dependent ferromagnetism $m(x)$, the Weyls and Fermi energy will move in tandem, $dE_\text{Weyl}/dn \approx dE_F/dn$. The shape of the AHC and ANC as a function of doping therefore only depend on the Weyl distance from the Fermi level, $\Delta E = E_\text{Weyl} - E_F$. Since the Weyl points remain a fixed distance from the top of the Fermi sea ($E_F$), we term this effect ``Fermi-surfing.'' 
%The combination of U-$5f$ kagome flat bands and doping-dependent magnetism in UCo$_{1-x}$Ru$_{x}$Al lead to this ``Fermi-surfing'' of the critically tilted Weyl points, which as we will show, results in the apparent violation of the Mott relation between $\sigma_{xy}$ and $\alpha_{xy}$. 

%How do Weyl points move as a function of tilt (no tilt = no motion!)
% For critically tilted Weyl points, the Weyls move exactly with the flat band
% The flat band itself creates a large density of states, meaning that as it shifts with changing magnetism, it stays fixed
% the combination of U-5f kagome flat bands and doping-dependent magnetism lead to this Fermi-surfing effect in the UCoRuAl system
% The flat bands create the critically tilted Weyl points AND the large density of states which pins the Fermi level to the flatband even as the system is doped. 

To make this argument more explicit, we can incorporate the magnetism and doping dependence to explain the behavior of $\sigma_{xy}$ and $\alpha_{xy}$.
%in UCo$_{1-x}$Ru$_{x}$Al and other magnetic TWM. 
For a single set of symmetry-equivalent Weyl points, the AHC can be approximated as a Gaussian \cite{noky_spectroscopy}, $\sigma_{xy} \propto e^{-(\varepsilon(n) - \varepsilon_p)^2/k_B T}$, which peaks around some critical value $\varepsilon_p$. In an ideal Weyl semimetal with no other topological features, this peak will occur at the energy of the Weyl points $\varepsilon_p = \varepsilon_\text{Weyl}$. The ANC will then have the two peak structure expected from the Mott relationship:  $\alpha_{xy} \propto (\varepsilon(n) - \varepsilon_p) e^{-(\varepsilon(n) - \varepsilon_p)^2/k_B T}$. 
We can approximate the energy relative to the peak at $\varepsilon_p$ 
%Based on Fig. \ref{eng_and_model}a, we can approximate the energy dependence 
as parabolic $(\varepsilon(n) - \varepsilon_p) \sim (n-n_0)^2$, 
%which incorporates magnetic and doping effects. 
which mimics the Fermi level moving towards and then away from the Weyl points with changing doping and magnetism.
Inserting this relation into our expressions for  $\sigma_{xy}$ and $\alpha_{xy}$, we see that both quantities will peak around the same doping, $n=n_0$. Furthermore, due to the Fermi surfing effect, there is an extended region near $n=n_0$ where $(\varepsilon(n) - \varepsilon_p) \sim \text{const.}$, which broadens the region in which $\sigma_{xy}$ and $\alpha_{xy}$ are maximal. 

%%%%%%%%%%%%%%%%%%%%
% RESULTS TB MODEL %
%%%%%%%%%%%%%%%%%%%%

%\begin{figure}[t]
	%\captionsetup{justification=raggedright}
%	\includegraphics[width=1.0\columnwidth]{fig2_eng_vs_nfill_v3.pdf}
%	\caption{Self-consistent Fermi energy (blue), Weyl point energy (green), and magnetization (red) vs. the electron filling. Red circle marks doping at which Weyl points appear. Gray box near half-filling ($n=3.0$) indicates the region where $E_F$ is pinned to the flat band forming the Weyl points.}
%	\label{energies}
%\end{figure}

\begin{figure*}[t]
	%\captionsetup{justification=raggedright}
	\includegraphics[width=1.0\textwidth]{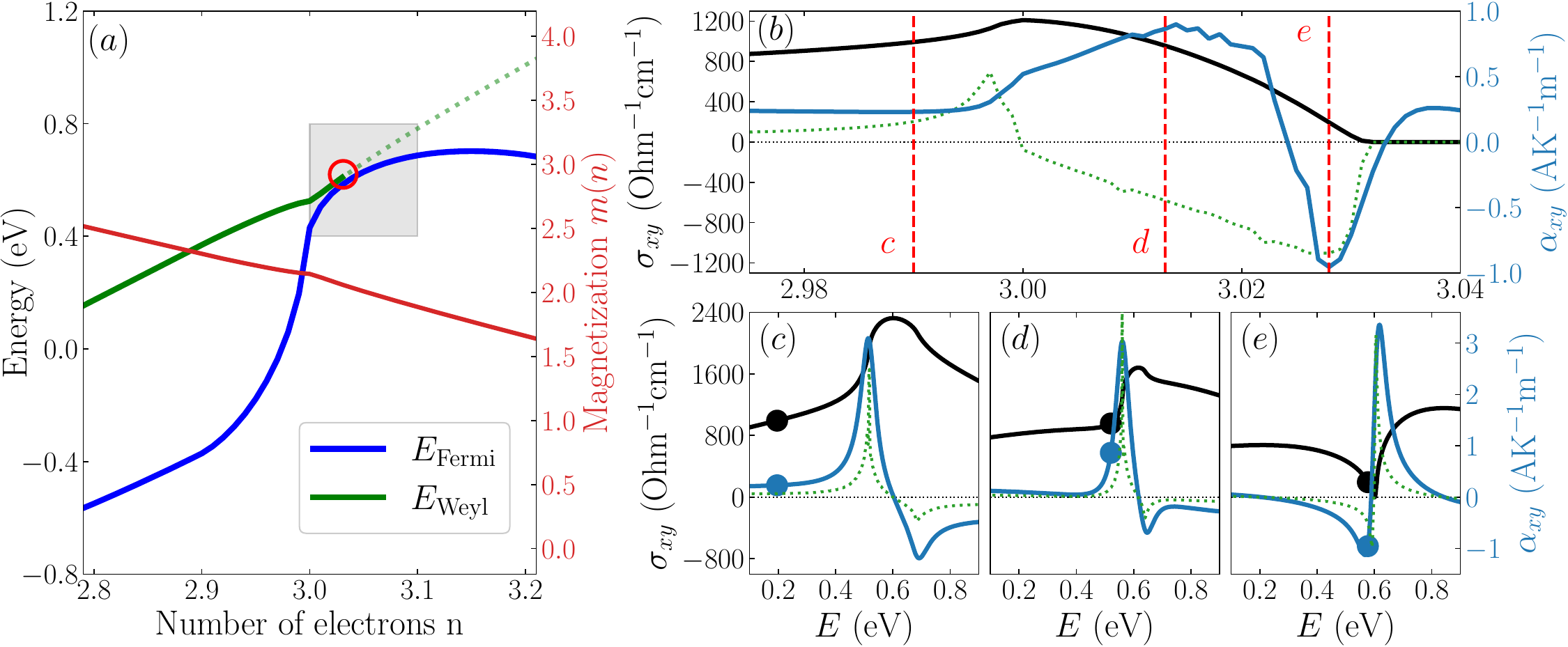}
	\caption{}
	\caption{(a)Self-consistent Fermi energy (blue), Weyl point energy (green), and magnetization (red) vs. the electron filling. Red circle marks doping at which Weyl points appear. Gray box near half-filling ($n=3.0$) indicates the region where $E_F$ is pinned to the flat band forming the Weyl points. (b) Computed anomalous Hall $\sigma_{xy}$ (black) and Nernst $\alpha_{xy}$ (blue) coefficients vs. doping, compared against $\alpha_{xy}$ expected from the Mott relation (green dashed). Red dashed lines mark specific dopings (c,d,e) for which $\sigma_{xy}$ (black), $\alpha_{xy}$ (blue), and $\sigma^\prime_{xy}$ (green dashed) are plotted vs. energy. Circles mark the values of $\sigma_{xy}(E_F)$ and $\alpha_{xy}(E_F)$ at the Fermi level.}
	\label{eng_and_model}
\end{figure*}

We now proceed to describing the critical Type-1.5 Weyl points in UCo$_{1-x}$Ru$_{x}$Al using a three band tight-binding model with Hubbard interaction treated at the mean-field level (see Methods).
Figure \ref{eng_and_model}a shows the behavior of the Fermi level and the energy of the Weyl points as a function of the electron filling. As the filling decreases, the Weyl points first appear around $n \sim 3.027$, and move to lower energy as the band splitting increases with increasing magnetization of the system. Concurrently, the Fermi level remains relatively constant, becoming pinned to the flat band hosting the Weyls between $n=3.1$ and $n=3.0$, before rapidly dropping due to a band becoming fully unoccupied as the system passes through $n=3.0$ half filling. As we will show below, the tandem motion of the Weyl points and the Fermi energy in the gray region in Fig. \ref{eng_and_model}a, followed by the rapid motion of $E_F$ away from the Weyls, fully determines the shape of the AHE and ANE as a function of doping, rather than the Mott relationship between them.

%%%%%%%%%%%%%%%%%%%%
% RESULTS: AHE/ANE %
%%%%%%%%%%%%%%%%%%%%

%We now move to the calculation of AHE and ANE (write formulas, cite literature and wanniertools)
%describe integration grid for Wanniertools
%refer to alternative model, say that additional sources of Berry curvature in the BZ lead to an extra peak, though the overall behavior is the same, and the Mott relation appears violated.
The AHE and ANE computed at 10K using the tight-binding model are shown in Figure \ref{eng_and_model}b, scaled by a factor of three to account for the three critically tilted Weyl pairs identified in UCo$_{1-x}$Ru$_{x}$Al \cite{ucra-sciadv}. As a function of doping, both quantities peak around $n \sim 3.01$, in the middle of the ``Fermi-surfing'' region, appearing to violate the Mott relation just as in the experimental measurement. However, plotting as a function of energy (Fig. \ref{eng_and_model}c,d,e) reveals good agreement between $\alpha_{xy}$ and the energy derivative of $\sigma_{xy}$ (green dashed lines), though the latter is narrower due to the lack of a temperature broadening parameter. Therefore, the Mott relation indeed holds at each doping independently. As the filling is decreased, net magnetization $m$ rises (Fig. \ref{eng_and_model}a), leading to a gradual increase in magnitude of $\sigma_{xy}$. $\alpha_{xy}$ also increases in magnitude with the magnetization, though the relationship is less straightforward, since the increasing steepness of $\sigma_{xy}$ also has a substantial effect. 

%It is clear that while the position of the Weyl points (and hence the peak maxima of $\sigma_{xy}$ and $\alpha_{xy}$) shifts with doping, the value measured at the Fermi energy is entirely determined by its relative position to the Weyls. 

\begin{figure}[b]
	%\captionsetup{justification=raggedright}
	\includegraphics[width=0.5\textwidth]{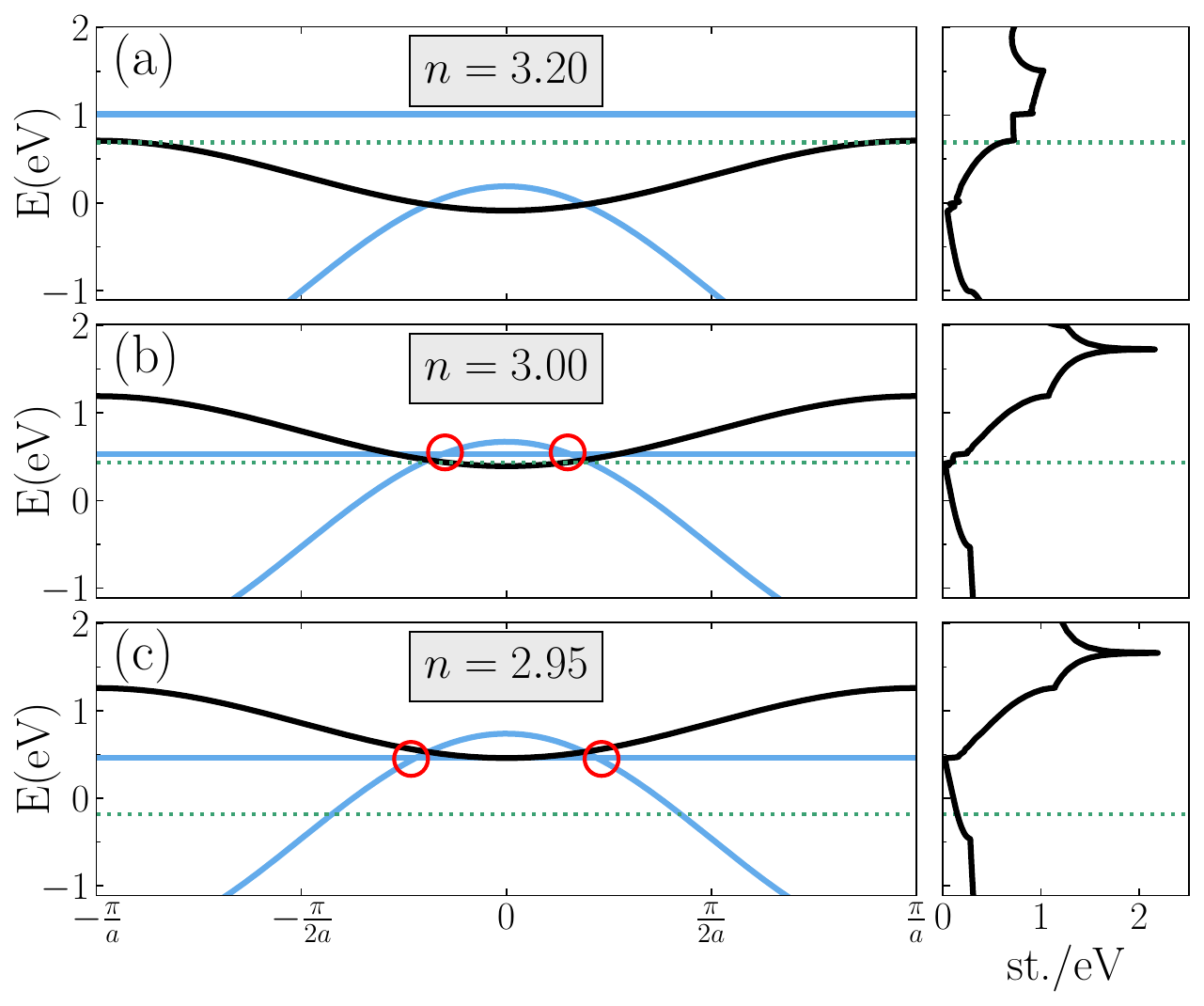}
	\caption[Bandstructure and Density of States for TB Model]{Self-consistent bands  (right)\& density of states (left) for the model in the text at (a) $n=3.2$, (b) $n=3.0$, and (c) $n=2.95$. The trivial band (black) crosses two bands (blue) forming the Weyl points (circled) above $E_F$ (dashed line).}
	\label{bands}
\end{figure}

The relationship of the Weyls and Fermi level to the doping behavior of $\sigma_{xy}$ and $\alpha_{xy}$ can be understood from the bandstructure and density of states (DOS) shown in Figure \ref{bands}. At large $n$, the Fermi level lies within a peak in the DOS. Therefore, $E_F$ remains approximately fixed as $n$ decreases, increasing magnetization $m(n)$ and causing the flat band to move towards $E_F$, leading to a concomitant increase in $\sigma_{xy}$ and $\alpha_{xy}$. As doping continues to increase and the flat band moves downward in energy, a large peak in the DOS emerges just above half-filling ($n=3.0$), keeping $E_F$ pinned to its energy. Here, in the so-called ``Fermi-surfing'' region, the Weyl points and $E_F$ move together, leading $\sigma_{xy}$ and $\alpha_{xy}$ to peak and remain roughly constant. Below half-filling, the Fermi level enters a region of low DOS, moving downward rapidly as $n$ is decreased further. As $E_F$ moves away from the Weyl points, $\sigma_{xy}$ and $\alpha_{xy}$ decrease. Therefore, the doping-dependent behavior of $\sigma_{xy}$ and $\alpha_{xy}$ is divorced from the Mott relation and is governed entirely by the motion of $E_F$ relative to the Weyls.

%%%%%%%%%%%%%%%%%%%%
%    DISCUSSION    %
%%%%%%%%%%%%%%%%%%%%

\section{Discussion}

The computed values of $\sigma_{xy}$ and $\alpha_{xy}$ from the model qualitatively match the experiment as shown in Fig. \ref{expt}. There is also good quantitative agreement with the AHE peak value, although the ANE peak is lower than the experimental value, which which reaches a maximum of $\sim$2.4 A K$^{-1}$ m$^{-1}$ at 10 K. The location of the maximum is sensitive to the simulation temperature, density of states, and the Weyl point positions, so the discrepancy may be due to the computed electronic structure and Weyl points 26meV above $E_F$ \cite{ucra-sciadv} being different in the real material.
%The peak of the ANE, however, is substantially smaller than what is seen in experiment. The reason for this is twofold. Firstly, as seen in Fig. \ref{eng_and_model}d, the Fermi level never aligns with the peak of the ANE, which depends on the simulation temperature and the energy of the Weyl points above $E_F$.
This may indicate a strategy to further enhance the colossal ANE of UCo$_{1-x}$Ru$_{x}$Al by tuning its magnetism (for instance, using pressure) simultaneously with doping, to control the energy of the Weyls relative to $E_F$. Alternatively, if a suitable dopant can be identified, doping-induced strain can be used instead. Secondly, the tilting of the Weyl points found by DFT is significantly closer to criticality, which would lead to a divergence of the ANE. However, simulating such tiltings would require prohibitively large $k$-point grids to converge $\alpha_{xy}$. In addition to the difference in magnitude, $\alpha_{xy}$ also has an additional negative peak feature between $n=3.025$ and $n=3.04$, which is due to the presence of other sources of Berry curvature in the model besides the Weyl points. Notably, there is also an order of magnitude discrepancy in the doping scale between the model and experiment. 
This is due to the much larger density of states ($10$-$30$ st./eV) in the DFT calculation, due to the U-$5f$ states near $E_F$ \cite{supplement}. This larger density of states would lead to a more sudden onset of the magnetism, with both $\sigma_{xy}$ and $\alpha_{xy}$ being necessarily zero above the critical region, which may result in features above $n=3.025$ not being visible to experiment.

Incorporating the effects of magnetism can capture the complex doping dependence of transport properties in other TWM. We will illustrate this using the case of Co$_3$Sn$_2$S$_2$, a well studied ferromagnetic Weyl semimetal material exhibiting a modest anomalous Hall and a large anomalous Nernst coefficient. The ferromagnetism in this TWM decreases monotonically upon either Ni-doping (ntype) \cite{Thakur2020_co3sn2s2_ni_doping, Liu_2023_co3sn2s2_both_doping, 2020Shen_co3sn2s2_ni_doping} or Fe-doping (p-type) \cite{2020_Shen_co3sn2s2_fe_doped}. While first-principles calculations predict a peak in $\sigma_{xy}$ just below the Fermi energy \cite{2019Ding_co3sn2s2_disorder_theory}, experimental measurements see a peak in $\sigma_{xy}$ both for n-type (Ni) \cite{2020Shen_co3sn2s2_ni_doping} and p-type (Fe) \cite{2020_Shen_co3sn2s2_fe_doped} doping. Once again, we approximate $\sigma_{xy}$ as a Gaussian \cite{noky_spectroscopy, 2019Ding_co3sn2s2_disorder_theory}, noting that Fe/Ni doping changes the Fermi level approximately linearly, $\varepsilon(n) = A n$. The spin splitting of the bands forming the Weyl points is reduced with decreasing ferromagnetism \cite{rossi, 2021_Liu_co3sn2s2_topotransition, 2021_Belopolski_co3sn2s2_topotransition}, causing the Weyl points, and associated peak in $\sigma_{xy}$ to move to higher energies. We approximate this behavior as $\varepsilon_c \sim \varepsilon_0 + B |n|$, where $\varepsilon_0<0$ represents the position of the $\sigma_{xy}$ peak maximum below $E_F$ in undoped Co$_3$Sn$_2$S$_2$ and $B>A>0$ captures the rapid motion of the Weyl points upward in energy relative to the motion of $E_F$ as Co$_3$Sn$_2$S$_2$ is doped. Using these expressions, we find that $\sigma_{xy}$ exhibits two peaks $n_\text{peak} = \pm \varepsilon_0/(B\pm A)$, for both electron and hole doping, just as seen in experiment. While this demonstrates the effectiveness of this qualitative argument for understanding complex behaviors of transport properties in TWM that include the effects of changing magnetism, doping, and energies, we note that a complete picture would need to also consider correlations \cite{rossi} and disorder \cite{2019Ding_co3sn2s2_disorder_theory, 2020Shen_co3sn2s2_ni_doping} effects, which have a substantial effect on the electronic properties of Co$_3$Sn$_2$S$_2$.

%%%%%%%%%%%%%%%%%%%%
%   CONCLUSIONS    %
%%%%%%%%%%%%%%%%%%%%

\section{Conclusion}

%In this Letter, 
We have reported doping dependent measurements of the AHE and ANE in UCo$_{1-x}$Ru$_{x}$Al, which appear to exhibit an apparent violation of the Mott relation. Using a suitable model for the critically tilted Weyl points predicted by DFT for this system, we show that this apparent violation results due to a ``Fermi surfing'' effect, in which the Fermi level is pinned to the flat band forming the Weyls because of its large density of states. With decreasing doping, the Fermi level approaches the AHE/ANE peaks, sampling progressively higher values, before rushing away as the band becomes fully unoccupied. In this way, the observed AHE/ANE are influenced not only by the energy dependence of $\sigma_{xy}$ and $\alpha_{xy}$, but also by the doping dependence of the magnetism and the density of states, which can be interpreted as a failure of the rigid-band approximation.

While the discussion presented here focuses primarily on explaining the unusual behavior of AHE/ANE in UCo$_{1-x}$Ru$_{x}$Al, these results have wider significance. An important consequence is for efforts that intend to use doping dependent measurements of the AHE and ANE as a spectroscopic method to track the energies of Weyl points in TWM. In magnetic and correlated materials, the non-uniform motion of the Fermi level due to variations in the density of states, along with the doping dependence of the magnetism, is needed to properly interpret transport measurements. As we have demonstrated for UCo$_{1-x}$Ru$_{x}$Al, a tight binding model that properly incorporates Hubbard interactions to match the magnetism and Fermi level with doping can correctly reproduce doping-dependent topological transport properties, even for doping with a magnetic element.

More generally, such careful treatment of the magnetic and density of states effects can enable the design of materials with simultaneously large AHE and ANE signals across a wide range of dopings, in contrast to prior proposals that focused on optimizing these at a single doping \cite{felserTaAs, 2025_ivanov_coactiveWeyls}. Furthermore, it is clear that the peak in the anomalous Hall and Nernst persists so long as the Fermi energy is located near the band forming the Weyl points. Therefore, in a material where the Weyl points are formed by a flat band with high density of states (such as in Kondo Weyl semimetals \cite{weyl_kondo1, weyl_kondo2}), a large AHE and ANE can be engineered to be doping resistant. This could be particularly useful for applications where electronic or thermal components are exposed to ionizing radiation. Such radiation would create vacancies, leading to disorder and hole doping of the material. While it is known that topological transport properties such as ANE can be resistant to or even enhanced by disorder \cite{nernst_disorder}, this can be combined with an engineered ``Fermi surfing'' region to make these materials tolerant to the electronic doping as well.

\section{Methods}

Polycrystalline samples of UCo$_{1-x}$Ru$_{x}$Al with Ru concentrations $x = 2.5\% - 70\%$ were synthesized by arc-melting the constituent elements on a water-cooled copper hearth using a Zr getter.  The arc-melted buttons were flipped four times to improve homogeneity.  The samples were wrapped in Ta foil, sealed in a silica ampoule under vacuum, and annealed at 900 $^\circ$C for 2 weeks. Each composition was confirmed to have the ZrNiAl-type structure by x-ray diffraction, and magnetic properties consistent with prior work \cite{Andreev1996_onset_of_ferromagnetism}. Thermoelectric and electronic transport measurements were performed using a single combined setup (Figure \ref{overview}). Thermocouples were used to measure the temperature gradient, which was generated by attaching one end of each sample to a 10 k$\Omega$ chip resistor, and the other end to a sapphire substrate clamped to a copper cold finger. Further details can be found in the supplemental material \cite{supplement}.

The behavior of the critical Type-1.5 Weyl points in UCo$_{1-x}$Ru$_{x}$Al is described using a three band tight-binding model with Hubbard interaction treated at the mean-field level. This mean-field approach, while computationally straightforward, is sufficient to model the metamagnetic physics of this material \cite{Yamase_2023_metamagnet_mf_model}. 
In momentum space, 
\begin{equation}
	\mathcal{H}(k) = 
	\begin{bmatrix}
		A_{+} + U\frac{m}{2}\sigma_z & \Gamma_{so} & 0 \\
		\Gamma_{so} & A_{-} + U\frac{m}{2}\sigma_z & 0 \\
		0 & 0 & C +  U\frac{m}{2}\sigma_z
	\end{bmatrix},
\end{equation}
where the bands are given by $A_{\pm}(k) = \pm \varepsilon \mp 2t \sum_{\xi} \cos k_{\xi} + \gamma \cos k_z$ and $C(k) = \varepsilon_c - 2t_c \sum_{\xi} \cos k_{\xi}$, while term $\Gamma_{so} = \sum_{\xi} t_{so}^\xi \sin k_{\xi} \sigma_\xi$ generates a spin-orbit splitting, with $\sigma_\xi \in \{\sigma_x, \sigma_y, \sigma_z\}$ being the Pauli matrices. The block of $A_\pm$ bands is very similar to commonly-used minimal TWM models \cite{rauch, noky_spectroscopy} with $\gamma$ controlling the Weyl cone tilt in the $k_z$-direction, while the $C(k)$ generates a non-vanishing density of states $g_\alpha(E)$ at the Weyl energy necessary for a metallic Weyl system. The magnetic splitting of the bands is determined by the Hubbard $U$ parameter and the magnetization $m =  \int_{-\infty}^{E_F} dE \left[g_\downarrow(E) - g_\uparrow(E)\right]$, which is determined self-consistently by tetrahedron integration on a $31\times 31 \times 31$ $\bm{k}$-point grid \cite{tetrahedron}. 
This model can readily be diagonalized along the $k_z$ direction to yield the energy $E_\text{Weyl} = \gamma \left[1+(m+\varepsilon)/2t \right]$ and position $k_\text{Weyl} = \arccos \left[(t+\varepsilon+m)/t\right]$ of the Weyl points. Notably, the tilting parameter $\gamma$ needs to be non-zero for the Weyl to move in energy; for critically tilted \cite{Sakai2018} Weyl points $\gamma=2t$.
More details of the tight-binding model are provided in the Supplementary Material \cite{supplement}.

The parameters used for the model are as follows: a $U=4.0$ eV Hubbard parameter was chosen to generate magnetic order, along with a spin-orbit coupling of $t_{so}^\xi = \{0.5,0.5,0\}$ eV necessary to create the Weyl points. The remaining parameters, $\varepsilon = -0.27$ eV, $\varepsilon_c = -1.75$ eV, $t = 0.3$ eV, $t_c = 0.2$ eV $\gamma = 2t-10^{-3}$ eV  were selected to match the narrow 0.2-0.3 eV U-$5f$ electronic bands  \cite{supplement}, and reproduce the energy  ($E_F + 26$ meV) and separation (0.06 1/\AA) of the Type-1.5 Weyl points identified by the DFT calculation.
%The parameters for the model are chosen as follows: $U=4.0$ eV, $\varepsilon = -0.27$ eV, $t = 0.3$ eV, $t_{so}^\xi = \{0.5,0.5,0\} eV$, and $\gamma = 2t-10^{-3}$ eV to tilt the Weyl points close to the transition between Type-I and Type-II. In order to mimic the large density of states in UCo$_{1-x}$Ru$_{x}$Al, an additional narrow band with $t_c = 0.2$ eV was introduced \cite{supplement}. When this band is placed at $\varepsilon_c = -1.75$ eV, at a filling $n = 3.0262$ the Weyl crossings achieve a minimal energy of 26 meV above the Fermi level, and a separation of 0.06 1/\AA, in line with prior DFT simulations \cite{lmtart, ucra-sciadv}. 
In the supplementary material, we provide an alternative set of parameters that produces a greater separation of the Weyl points, and a more clear representation of the apparent violation of the Mott relationship between the AHE and ANE \cite{supplement}.

The anomalous Hall coefficient $\sigma_{xy}$ and the anomalous Nernst coefficient $\alpha_{xy}$ are computed using the WannierTools package \cite{wanniertools} at $T=10$K as weighted integrals of the Berry curvature $\Omega_{ij}(\bm{k})$ over the Brillouin zone:
\begin{align}
	\sigma_{ij} &= \frac{e^2}{\hbar} \int \frac{d\bm{k}}{(2\pi)^{3}} \Omega_{ij}(\bm{k}) f_{\text{FD}}(\bm{k}) \nonumber\\
	\alpha_{ij} &= \frac{e k_B}{\hbar} \int \frac{d\bm{k}}{(2\pi)^{3}} \Omega_{ij}(\bm{k}) s(\bm{k}),
	\label{ahe_ane_eq}
\end{align}
where $f_\text{FD}(\bm{k})$ is the Fermi-Dirac distribution, and $s(\bm{k}) = \left[\ln(1 + e^{-\beta(\epsilon-\mu)}) +  \beta(\epsilon-\mu) f_{\text{FD}}(\bm{k}) \right]$ \cite{Xiao2006_nernst}. Near the quantum Lifshitz transition, the Berry curvature near the Weyl points begins to diverge \cite{Sakai2018}, requiring a dense $600^3$ $\bm{k}$-point integration grid for convergence.
%where $\beta = 1/(k_B T)$, $\Omega_{ij}(\bm{k})$ is the Berry curvature. T 

%point out that since the Fermi level is pinned to the Weyl points , a similar effect in another strongly correlated system could enable maintained thermoelectric performance across a wide range of doping levels. 
% This is particularly relevant as it has been proposed that such nernst materials might be used for radioisotope thermoelectric generators - our work suggests that this "Fermi surfing" effect might mean that the performance is maintained even as the materials Fermi level drops due to the formation of vacancies
% However a more detailed study of the effect of disorder/material damage on alpha_xy itself is needed

%\section{\textbf{Acknowledgements}}
\begin{acknowledgments}
	VI acknowledges support from Virginia Tech startup funds. The authors acknowledge Advanced Research Computing at Virginia Tech (arc.vt.edu) for providing computational resources and technical support. The experimental work was performed at Los Alamos National Laboratory under the auspices of the U.S. Department of Energy, Office of Basic Energy Sciences, Division of Materials Science and Engineering. S.S acknowledges support from US DOE grant No. DE-SC0026106. 
\end{acknowledgments}

\section*{Author Contributions}

Benjamin Kostroun and Vsevolod Ivanov performed the theoretical calculations with input from  Sergey Y. Savrasov. Eric D. Bauer synthesized the UCo$_{1-x}$Ru$_{x}$Al samples. Tomoya Asaba, Sean M. Thomas, and Filip Ronning created the measurement setup and performed the transport measurements. Tomoya Asaba, Filip Ronning, and Vsevolod Ivanov performed the experimental data analysis. Filip Ronning and Sergey Y. Savrasov provided overall supervision of the project and acquired the necessary funding. Benjamin Kostroun and Vsevolod Ivanov drafted the manuscript with input from all other authors. The work was conceived by Filip Ronning, Tomoya Asaba, and Vsevolod Ivanov.

\section*{Competing Interests}
The authors declare no competing interests.

\section*{Data Availability}
The data that support the findings of this study are available from the corresponding author upon reasonable request.
\section*{Code Availability}
The code used to produce the results is available from the corresponding author upon reasonable request.

\bibliography{references}

\end{document}